\documentclass[cits]{PoS}
\usepackage{epsfig}

\title{The rooted staggered determinant in the Schwinger model}

\ShortTitle{The rooted staggered determinant in the Schwinger model}

\author{Anna Hasenfratz and \speaker{Roland Hoffmann}\\
        Departmen of Physics, University of Colorado, Boulder, CO-80309-390\\
        E-mail: \email{hoffmann@pizero.colorado.edu}}

\abstract{We investigate the continuum limit of the rooted staggered action
in the 2-dimensional Schwinger model. We match both the unrooted
and rooted staggered determinants with an overlap fermion determinant of two
(one) flavors and a local pure gauge effective action by fitting the coefficients of
the effective action and the mass of the overlap operator. The residue of this fit
measures the difference of the staggered and overlap fermion actions. We show
that this residue scales at least as ${\rm O}(a^2)$, implying that any difference,
be it local or non-local, between the staggered and overlap actions becomes irrelevant in 
the continuum limit. For the model under consideration here, this observation 
justifies the rooting procedure for the staggered sea quark action.}

\FullConference{XXIVth International Symposium on Lattice Field Theory\\
                July 23-28, 2006\\
                Tucson, Arizona, USA}

\begin{document}

\section{Introduction}

While staggered fermions offer many computational advantages,
their action does not have full chiral symmetry and the chiral limit has to
be taken together with the continuum limit. This is no different from other non-chiral actions,
but staggered fermions have another, potentially serious problem.
In 4 dimensions the staggered action describes four species (or tastes)
of fermions and in order to reduce the number of tastes a fractional power
of the fermion determinant is taken in the path integral. There
is no a priori reason that this rooted determinant corresponds to
a local fermionic action belonging to the same universality class
as 1--flavor QCD.

Although several analytical and numerical works addressed this question in
the last few years \cite{Bunk:2004br,Adams:2004mf,Maresca:2004me,Shamir:2004zc,Durr:2003xs,Neuberger:2004be,Durr:2004ta,Bernard:2004ab,Bernard:2005gf,Hasenfratz:2005ri},
none of them showed evidence that the procedure introduces non--universal
errors, i.e. errors that cannot be considered cutoff effects. Recently it has
been argued, based
on a number of reasonable conjectures, that while the rooted staggered
action is non-local at any finite lattice spacing, in the continuum
limit the non-local terms become irrelevant \cite{Bernard:2006ee,Shamir:2006nj}.

In this paper we present numerical evidence obtained in the massive
Schwinger model, showing that the rooted staggered action is in the
right universality class. We
also show that the staggered action can be considered equivalent to
a chiral Ginsparg-Wilson action only when the staggered mass is larger
than typical taste symmetry breaking effects, limiting the parameter
space where staggered simulations can be expected to approximate continuum
QCD. We describe how the masses of the staggered and corresponding
overlap actions should be matched to obtain physically equivalent
theories when this condition is satisfied.
More details on the matching procedure and additional results
can be found in \cite{Hasenfratz:2006nw}.

\section{The continuum limit of the staggered action\label{sec:The-continuum-limit}}

The partition function of the unrooted staggered action is \begin{eqnarray}
Z & = & \int\!\!\! D[U\bar{\psi}\psi]\, e^{-S_{_{g}}(U)-\bar{\psi}(M+am_{{\rm {st}}})\psi}\label{Part.Func.} = \int\!\!\! D[U]\,{\rm {det}}(M+am_{{\rm {st}}})\, e^{-S_{g}(U)}\,, \end{eqnarray}
 where $S_{g}(U)$ is a gauge action, $M$ is the staggered Dirac
operator and $am_{{\rm {st}}}$ is the bare staggered mass. In
the $a\to0$ continuum limit the staggered action describes $n_{t}=4$
degenerate fermions in 4, $n_{t}=2$ fermions in 2 dimensions. At
finite lattice spacing the taste symmetry is broken, the action describes
$n_{t}$ fermion tastes but only with a remnant $U(1)$ taste symmetry.
Depending on the staggered quark mass, at finite lattice spacing
one has one of the following situations.
\begin{itemize}
\item At $am_{st}=0$ the staggered action's spectrum has a single Goldstone
particle and $n_{t}^{2}-1$ massive pseudoscalars. While $n_{t}^{2}-2$
of these will become massless as $a\to0$, at any finite lattice spacing
the staggered spectrum is very different from $n_{t}$--flavor massless
QCD. At small fermion mass $am_{{\rm {st}}}\gtrsim0$ the taste breaking
terms dominate and the non--Goldstone pions are heavy compared to
the Goldstone one. One does not expect QCD--like behavior.
\item $am_{{\rm {st}}}\gtrsim1$ is the cutoff region, again not continuum QCD--like.
\item Only in the middle of these extremes
would one expect to observe QCD. The $a\to0$, $am_{{\rm {st}}}\to0$
continuum limit should be approached here.
\end{itemize}
While staggered fermions formally allow $am_{{\rm {st}}}=0$, physically
this limit does not correspond to QCD at any finite lattice spacing
\cite{Durr:2004ta,Bernard:2004ab}. Simulations cannot be trusted
at a small fermion mass where taste breaking terms dominate the pseudoscalar
sector. However, the taste breaking terms are expected to scale at
least with $O(a^{2})$, such that at small enough lattice spacing
the continuum limit can be approached with any finite fermion mass.
Thus the exclusion of $am_{{\rm {st}}}=0$ is not a serious problem
for massive fermions.

The staggered determinant can always be written as
\vspace*{-2mm}\begin{equation}
{\rm det}(M+am_{{\rm {st}}})=\mathrm{det}{}^{n_{t}}(D_{{\rm {1f}}}+am_{{\rm {1f}}}){\rm \,{det}}(T)\,,\label{Det_relation}\vspace*{-2mm}\end{equation}
 where $D_{{\rm {1f}}}+am_{{\rm {1f}}}$ is an arbitrary 1--flavor
Dirac operator and ${\rm det}(T)$ describes all the terms that are
not included in the latter. If the local $D_{{\rm {1f}}}$ operator
and the mass term $m_{{\rm {1f}}}$ could be chosen such that $T$
contains only local gauge terms,\vspace*{-2mm}\begin{equation}
{\rm {det}}(T)=e^{-S_{{\rm {eff}}}(U)}\,\,,\label{local_T}\vspace*{-2mm}\end{equation}
 the staggered action would differ from an $n_{t}$--flavor degenerate
Dirac operator only in cutoff level terms \cite{Adams:2004mf}. This
is indeed the case for heavy, $am_{{\rm {st}}}\!\gtrsim\!1$ fermions.

On the other hand there are several examples \cite{Hasenfratz:2005ri,Bernard:2006ee,Kogut:2006gt}
that illustrate that
at $am_{{\rm {st}}}=0$ the operator $T$ cannot be local at any finite
lattice spacing. This, however, does not mean that the staggered operator
cannot describe QCD in the continuum limit. If we write the determinant
as \vspace*{-2mm}\begin{equation}
{\rm {det}}(T)=e^{-S_{{\rm eff}}(U)}\,{\rm {det}}(1\!+\!\Delta)\,,\label{non-local_T}\vspace*{-2mm}\end{equation}
and can choose $S_{{\rm {eff}}}$ such that the non--local term $\Delta$
is bounded at finite mass and goes to zero as $a\to0$, the staggered
determinant in Eq.(\ref{Det_relation}) will describe $n_{t}$ degenerate
flavors in the continuum limit. This is certainly the expected behavior
for the unrooted action.

Now we turn our attention to the rooting procedure. With the notation
introduced above, the root of the staggered determinant is
\vspace*{-2mm}\begin{equation}
\mathrm{det}^{1/n_{t}}(M+am_{{\rm {st}}})={\rm {det}}(D_{{\rm {1f}}}+am_{{\rm {1f}}})\, e^{-S_{{\rm eff}}(U)/n_{t}}\,{\mathrm{det}}^{1/n_{t}}(1\!+\!\Delta).\label{rooted_det}\vspace*{-2mm}\end{equation}
If one could show that\vspace*{-4mm}\begin{eqnarray}
\Delta\to0\,\, & {\rm {as}} & a\to0\,,\label{required}\vspace*{-2mm}\end{eqnarray}
 the rooted determinant of Eq.(\ref{rooted_det}) would correspond
to a local 1--flavor action in the continuum limit. 
Based on renormalization group arguments, in Ref. \cite{Shamir:2004zc}
Shamir showed that this is indeed the case for free fermions. In a
recent work \cite{Shamir:2006nj}, based on a number of reasonable
assumptions, he argues that the same is true in the interacting theory.
Here we present numerical results to support this
claim.

In the following we pick an arbitrary Ginsparg-Wilson  operator as
$D_{{\rm {1f}}}$ and ask if $am_{{\rm {1f}}}$ and $S_{{\rm {eff}}}(U)$
can be chosen such that Eq.(\ref{required}) is satisfied.

\section{Matching the fermionic determinants}

The actual matching strategy is fairly general and
we will describe it for an arbitrary pair of Dirac operators $D_{1}+am_{1}$
and $D_{2}+am_{2}$. We want to know to what extent the determinant
of the first Dirac operator can be described by the determinant of
the second plus pure gauge terms. To find this we calculate the determinant
ratio\vspace*{-2mm}
\begin{equation}
\mathrm{det}(T)=\frac{\mathrm{det}(D_{1}+am_{1})}{\mathrm{det}(D_{2}+am_{2})}\label{det_ratio}\end{equation}
on configurations generated
with the action $
S_{1}=S_{g}(U)+\bar{\psi}(D_{1}+am_{1})\psi\,.$
Next we fit the logarithm of the determinant ratios with a local pure gauge action
$S_{{\rm eff}}$. In practice
we use an ultralocal effective action consisting of small Wilson loops. The accuracy
of the matching at fixed fermion mass $m_{2}$ is characterized by
the per flavor/taste residue
\vspace*{-1mm}
\begin{equation}
r(m_{2})=\Big\langle\Big({\rm {log}\,}\frac{{\rm {det}}(D_{1}+am_{1})}{{\rm {det}}(D_{2}+am_{2})}-S_{{\rm {eff}}}(U)\Big)^{2}\Big\rangle^{1/2}\,.\label{residue}\vspace*{-1mm}\end{equation}
The minimum of the residue $r(m_{2})$ in terms of $m_{2}$ determines
the action $D_{2}+am_{2}$ that is \emph{physically
closest} to the original $D_{1}+am_{1}$ action.
In this sense it defines the mass $\overline{m}_{2}$ that matches
the fermion mass $m_{1}$. In the notation of Eq.(\ref{non-local_T})
then\vspace*{-2mm}
\begin{equation}
r(\overline{m}_{2})=\Big\langle\Big(({\rm {log\, det}}(1\!+\!\Delta)\Big)^{2}\Big\rangle^{1/2}\,.\label{matched_r}\vspace*{-2mm}\end{equation}
If the two fermion operators describe the same continuum theory the
residue has to vanish as $a\to0$ at fixed volume and quark mass.

\section{Schwinger model - numerical results \label{sec:Schwinger-model}}

\vspace*{-2mm}
\subsection{Setup and matching tests }

The 2--dimensional Schwinger model offers an excellent testing ground
for the matching idea as it can be studied with high accuracy and
limited computer resources. It is a super--renormalizable theory since
the bare gauge coupling $g$ is dimensional, the lattice gauge coupling
is $\beta=1/(ag)^{2}$. A continuum limit in fixed physical volume
can be achieved by keeping the scaling variable $z=Lg$ fixed while
increasing the lattice resolution. We choose $z=6$ and vary the lattice
size between $L/a=12$ and $L/a=28$. The scaling parameter $z$ characterizes
the (physical) volume while we use $mL$ to fix the mass. 

We produced gauge configurations using a global heatbath for the plaquette
gauge action and in the data analysis the measurements are
reweighted with the appropriate power of the fermion determinant to
obtain the observables in the full dynamical theory. On the gauge
configurations we measure a set of Wilson loops $\mathcal{C}_{l}$
as well as the complete spectra of the Dirac operators under consideration.
For the matching we use 9 loops up to length 10 in $S_{\rm eff}$. With a maximal extension of
four lattice units $S_{{\rm eff}}$ is very localized even on our
coarsest lattices and in particular we do \emph{not} increase the
size or number of loops as we approach the continuum.
In the following we concentrate on the matching of the staggered action
to a smeared overlap action. For more details see \cite{Hasenfratz:2006nw}.

\subsection{The unrooted staggered action}

We start our investigation with the unrooted action, which in 2 dimensions
corresponds to two fermion tastes. In the continuum limit it is expected
to describe two degenerate flavors and thus it should differ from
a degenerate 2--flavor overlap action in irrelevant terms only.
\DOUBLEFIGURE[t]{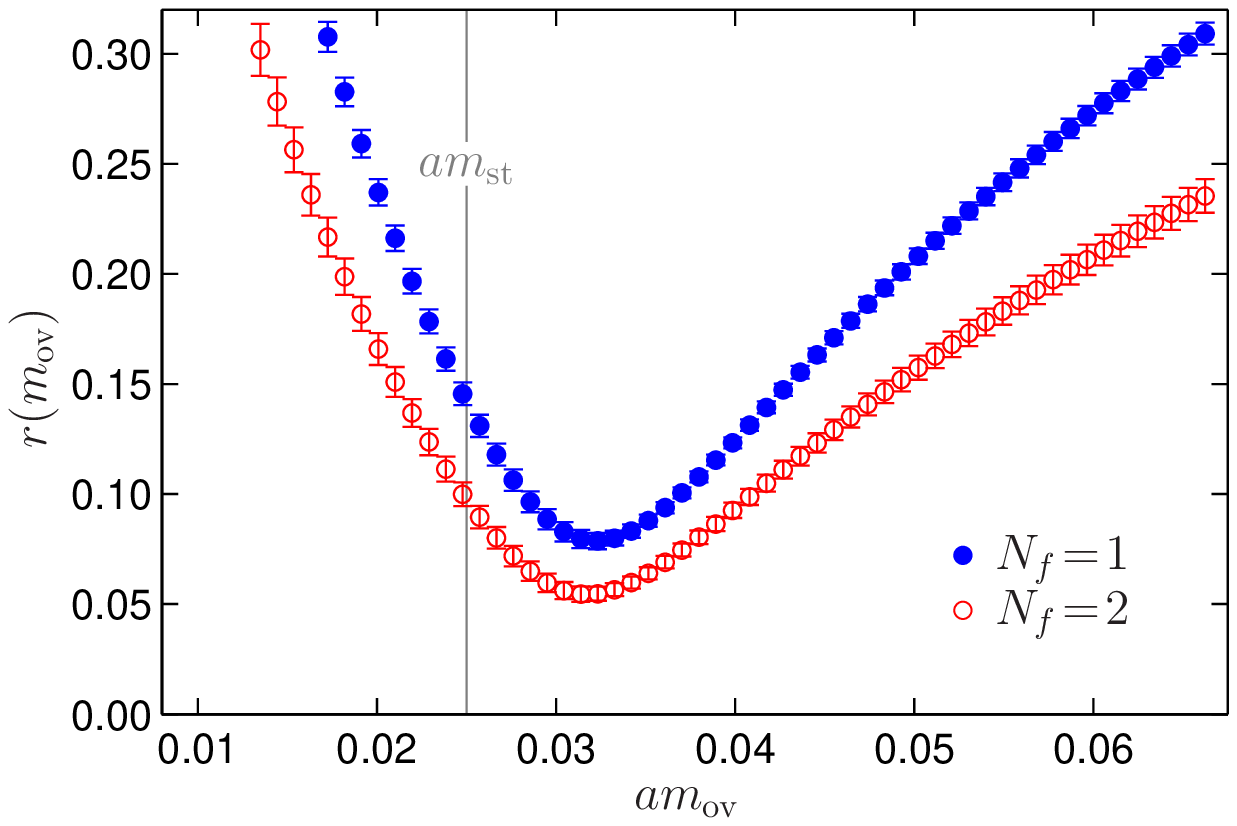,width=72mm}{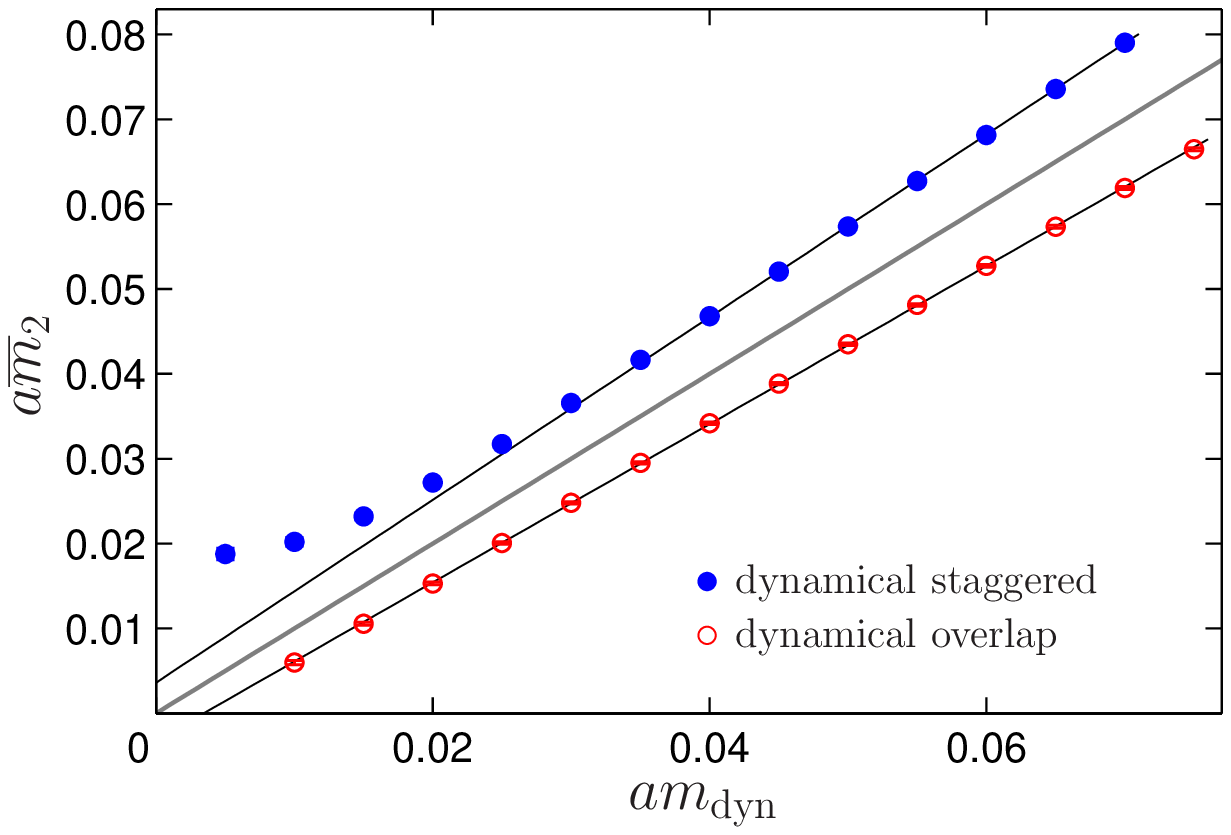,width=72mm}{
The residue of Eq.(\protect\ref{residue}) as the function of the matching
mass on $L/a=20$ configurations generated at $am_{{\rm {st}}}=0.025$.
\label{fig:minimum}
}
{
The matching mass as the function of the dynamical action mass
at fixed lattice spacing for 2 tastes/flavors.\label{fig:Chiral}
}
Fig.\ref{fig:minimum} shows the matching of the $n_{t}=2$ staggered
determinant with the $N_{f}=2$ flavor-degenerate overlap determinant
at $z=6$ on $L/a=20$ lattices ($\beta\simeq11.11$). The quenched
configurations were reweighted to the dynamical staggered ensemble
at $am_{{\rm {st}}}=0.025$. The residue of the matching (Eq.\ref{residue})
has a well defined minimum at $a\overline{m}_{{\rm ov}}=0.0317(3)$. 

By repeating the matching at different values of the staggered masses
$am_{{\rm {st}}}$ we can find the matching overlap masses at the
given lattice spacing as shown by the blue dots in Fig.\ref{fig:Chiral}.
For larger masses the data show a linear dependence with a constant
offset.
For small masses, below $am_{\mathrm{st}}\approx0.02$, there is a
clear deviation from the linear behavior. The residue of the fit increases
from 0.02 at the heaviest mass to 0.09 at the lightest one, indicating
that the matching is no longer meaningful. According to the discussion
in Sect.\ref{sec:The-continuum-limit} we interpret this as the staggered
action being QCD--like for $am_{\mathrm{st}}{\rm \gtrsim}0.02$ and
not QCD--like below. 
As a consistency check we repeated the same matching on a dynamical overlap
background. The result, shown by the red circles in Fig.\ref{fig:Chiral},
is the mirror image of the staggered with overlap matching data up
to the point where the latter matching breaks down. This is the expected
behavior if the two actions differ only by lattice artifacts.

By restricting the configurations to the sector of trivial topology
we could verify that the difference between the matching on the two
ensembles and also most of the residue can be ascribed to configurations
with non--vanishing topological charge.

Next we consider the continuum limit of the matching at fixed physical
mass
\footnote{We fix the physical mass by keeping $m_{{\rm ov}}L$ constant and
vary the staggered sea quark mass to achieve the matching.}
and volume $z=6$. Fig.\ref{fig:Residue}a shows the residue of
matching the 2-taste staggered determinant with the 2-flavor overlap
determinant at different masses as a function of $a^{2}g^{2}$.
 For the smallest mass, $m_{{\rm ov}}L=0.4$
the data stops around $a^{2}g^{2}=0.11$ - on coarser lattices the
two actions cannot be matched, the residue of Eq.(\ref{residue})
has no minimum. Nevertheless matching is possible at smaller lattice
spacing and the residue at fixed $m_{{\rm ov}}L$ approaches zero
at least quadratically in $a$. The continuum limit can be approached
with any fermion mass and the staggered determinant can be described
as a 2--flavor chiral determinant plus pure gauge terms. This is the
behavior we expected from universality.
\EPSFIGURE[t]{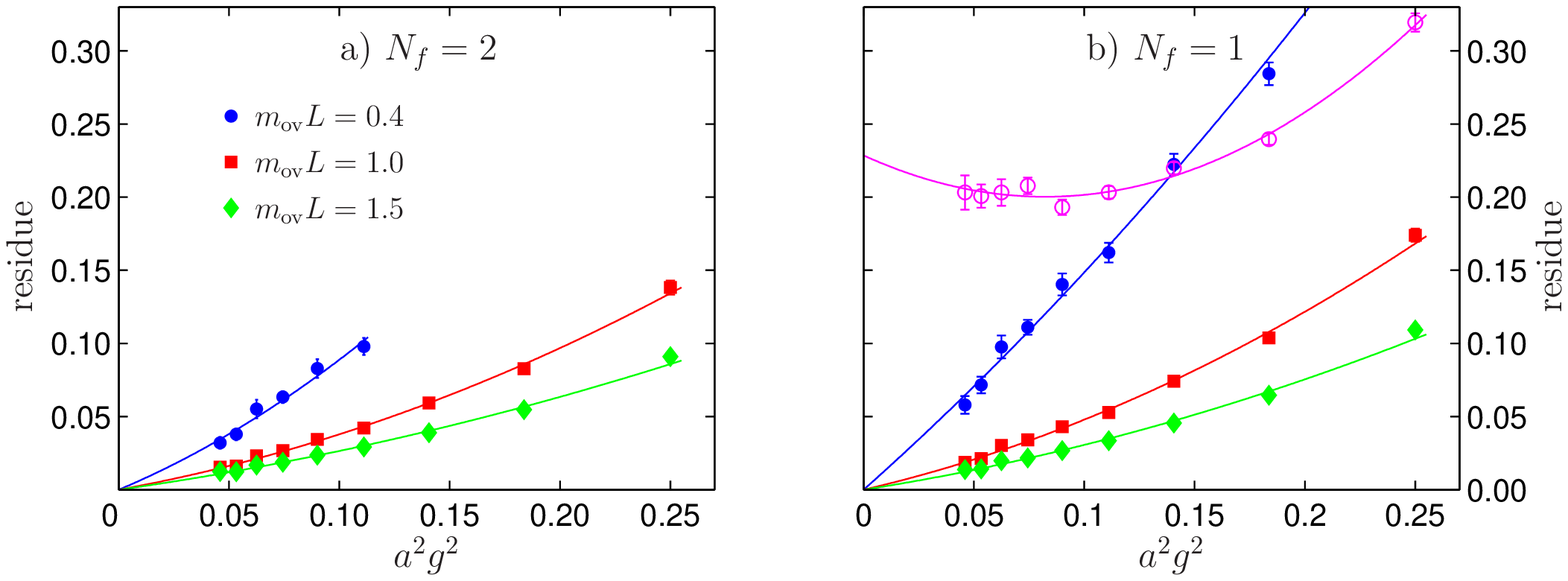,width=135mm}{
Residue of the matching as a function of the (squared) lattice spacing
at different physical masses. a) unrooted staggered/2--flavor overlap;
b) rooted staggered/1--flavor overlap matching. The open circles in
b) show the residue from matching a 1--flavor overlap ($m_{\mathrm{ov}}L=0.5$)
to an \emph{unrooted} staggered action.\label{fig:Residue}}

\subsection{The rooted staggered action}

Now we repeat the analysis of the previous section for the rooted
staggered action. The configurations
are reweighted with the square root of the staggered determinant
and the rooted determinant
is matched with the 1-flavor overlap determinant plus pure gauge terms,
according to Eq.(\ref{residue}).

The quality of the matching is very similar to the unrooted/2--flavor
case as the open circles in Fig.\ref{fig:minimum} show. In fact,
even the matched mass $a\overline{m}_{{\rm ov}}=0.0322(2)$ hardly
differs from the 2--flavor case. The 1--flavor data in Fig.\ref{fig:Chiral}
is indistinguishable from the shown 2--flavor data.

Fig.\ref{fig:Residue}b is the important plot for the rooted staggered
action as it shows the residue at fixed physical masses as the continuum
limit is approached. While the residue for the 1--flavor
rooted determinant is larger than in the unrooted case, the continuum
approach is identical,
at least quadratic in $a$. The taste violating term $\Delta$ in
Eqs.(\ref{non-local_T}) and (\ref{matched_r}) becomes irrelevant
in the continuum limit, a result that justifies the rooting procedure.
As a test of the sensitivity of our matching method we also show the
residue of an attempted matching of a 1--flavor overlap action with
an unrooted staggered action, which clearly does \emph{not} vanish
in the continuum limit.

\subsection{Application}

We can now apply our knowledge of the matching overlap mass in mixed action
simulations.
The first observable we consider is the topological susceptibility
$\langle Q^{2}\rangle/z^{2}$, as it
is very sensitive to the sea quarks. We define the topology
through the zero--modes of the smeared overlap operator used in the
matching and evaluate it on gauge ensembles generated with two and
one flavor/taste staggered and overlap actions at various masses.
This is the simplest case of a mixed action simulations as the observable
does not depend on the valance quark mass. Possibly more interesting
is the scalar condensate $\langle\bar\psi\psi\rangle$, which diverges in the limit of vanishing staggered mass due
to insufficient suppression of topologically non--trivial configurations \cite{Durr:2003xs}.

\DOUBLEFIGURE[t]{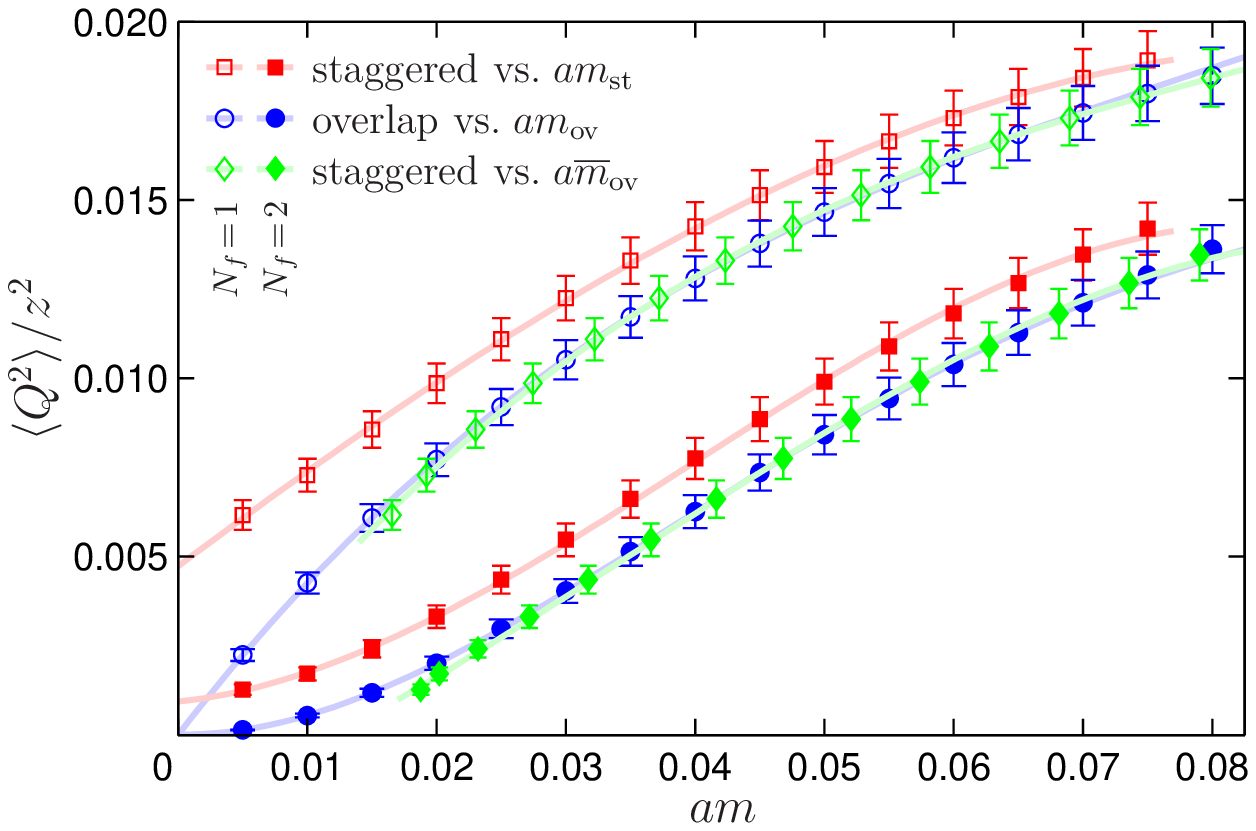,height=4.9cm}{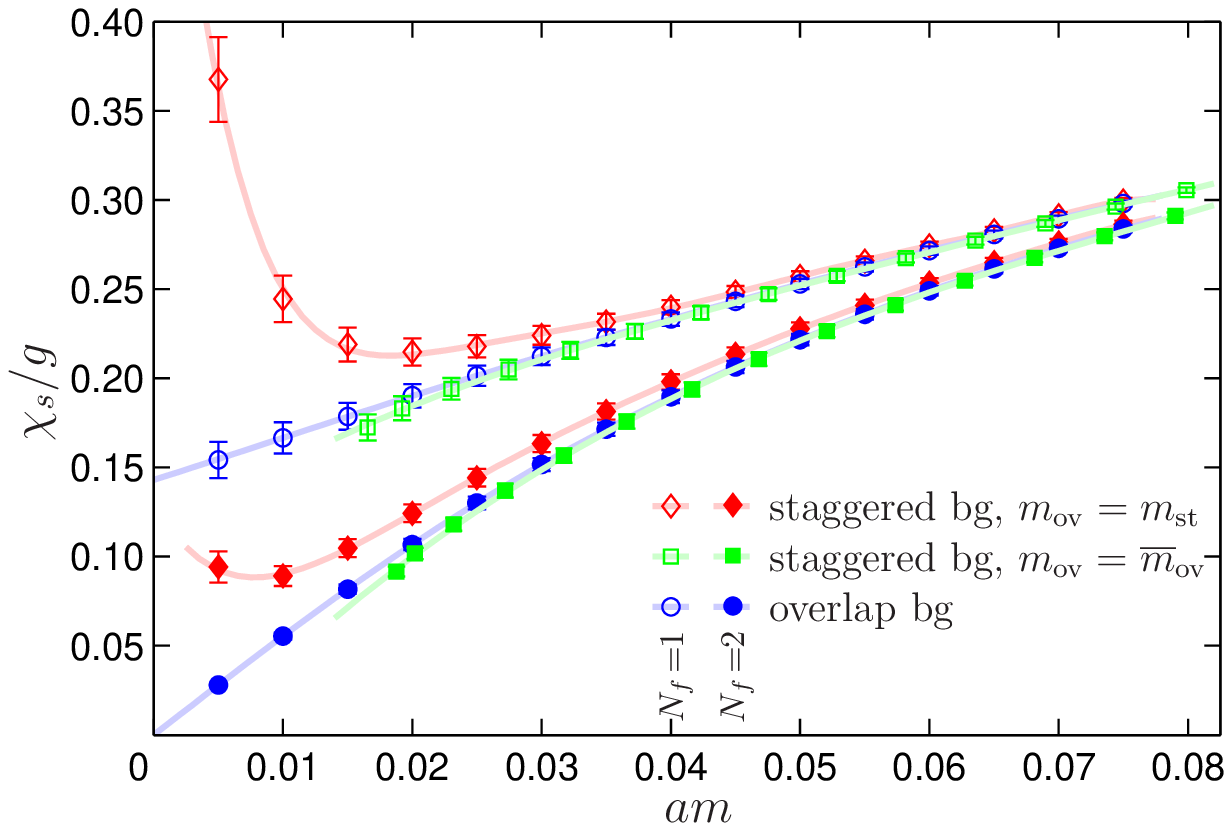,height=4.9cm}{
The topological susceptibility on the $L/a=20$ ensemble.\label{fig:topo}
After shifting the staggered data to the matched overlap
mass, almost perfect agreement with the overlap data
is achieved.}
{Also the scalar condensate $\chi_s$ on a staggered background agrees
with the overlap result when it is evaluated using the matched overlap mass.\label{fig:cond}
\label{final}}

Results from $L/a=20$ lattices are shown in Figs.\ref{fig:topo} and \ref{fig:cond},
where the difference of the staggered and overlap ensemble at the
same bare fermion mass is very evident, especially at small masses. After shifting the staggered data
to the matching overlap mass, excellent agreement
is achieved. One should note that the agreement for the 2 taste/flavor case, where the
difference between the actions is entirely due to local lattice artifacts, is
not better or significantly different than the rooted 1 taste/flavor case.
The agreement on our finer lattices is equally
good and extends to smaller quark masses.

\section{Conclusion}

The rooted staggered action is likely non--local in the physically
relevant range of small quark masses. However, this does not invalidate
the rooted action as long as the non--local terms are irrelevant and
scale away in the continuum limit. Here we demonstrated that this
is indeed the case in the 2--dimensional Schwinger model. We studied
how the staggered action differs from a chiral overlap action along
a line of constant physics as the continuum is approached. For both
the unrooted (as expected) and rooted staggered action we found that
the difference reduces to irrelevant operators plus local pure gauge terms.
Nevertheless
care is required in taking the continuum limit of staggered fermions
such that the non QCD--like region is avoided.

{\renewcommand{\baselinestretch}{0.95}
\bibliography{lattice}
\bibliographystyle{JHEP}}

\end{document}